\documentclass[twocolumn,showpacs,preprintnumbers,prb,amsmath,amssymb]{revtex4}

\usepackage{graphicx}
\usepackage{dcolumn}
\usepackage{bm}
\usepackage{color}

\begin{document}


\title{Effect of annealing on the specific heat of   Ba(Fe$_{1-x}$Co$_{x}$)$_{2}$As$_{2}$}

\author{K. Gofryk$^{1}$}
\email{gofryk@lanl.gov}
\author{A. B. Vorontsov$^{2}$}
\author{I. Vekhter$^{3}$}
\author{A. S. Sefat$^{4}$}
\author{T. Imai$^{5,6}$}
\author{E. D. Bauer$^{1}$}
\author{J. D. Thompson$^{1}$}
\author{F. Ronning$^{1}$}
\email{fronning@lanl.gov}
\affiliation{$^{1}$Condensed Matter and Thermal Physics, Los Alamos National Laboratory, Los Alamos, New Mexico 87545, USA\\$^{2}$Department of Physics, Montana State University, Bozeman, Montana, 59717, USA\\$^{3}$Department of Physics and Astronomy, Louisiana State University, Baton Rouge, Louisiana, 70803, USA\\
$^{4}$Materials Science and Technology Division, Oak Ridge National Laboratory, Oak Ridge, Tennessee 37831, USA\\$^{5}$Department of Physics and Astronomy, McMaster University, Hamilton, Ontario L8S4M1, Canada\\$^{6}$Canadian Institute for Advanced Research, Toronto, Ontario M5G1Z8, Canada}


\begin{abstract}

We report on the effect of annealing on the temperature and field dependencies of the low temperature specific heat of the electron-doped Ba(Fe$_{1-x}$Co$_{x}$)$_{2}$As$_{2}$ for under-(x = 0.045), optimal- (x = 0.08) and over-doped (x = 0.105 and 0.14) regimes. We observed that annealing significantly improves some superconducting characteristics in Ba(Fe$_{1-x}$Co$_{x}$)$_{2}$As$_{2}$. It considerably increases $T_{c}$, decreases $\gamma_{0}$ in the superconducting state and suppresses the Schottky-like contribution at very low temperatures. The improved sample quality allows for a better identification of the superconducting gap structure of these materials. We examine the effects of doping and annealing within a self-consistent framework for an extended $s$-wave pairing scenario. At optimal doping our data indicates the sample is fully gapped, while for both under and overdoped samples significant low-energy excitations possibly consistent with a nodal structure remain. The difference of sample quality offers a natural explanation for the variation in low temperature power laws observed by many techniques.

\end{abstract}

\pacs{74.20.Rp, 74.70.Dd, 74.62.Dh, 65.40.Ba}
\maketitle

\section{Introduction}

It is not surprising that two years after the initial discovery of superconductivity (SC) in Fe-based materials\cite{Kamihara,chen} there continues to be much debate as to the structure of the superconducting gap and the origin of the pairing mechanism. Even in the cuprates these questions are still not fully resolved, however significant progress was achieved with improved crystal quality. While the highest $T_{c}$ is found in the 1111 family of compounds\cite{ren}, more progress on the gap structure has been made on the 122 family of compounds\cite{122a,122b}, as large single crystals can be synthesized. While much has already been learned on these crystals\cite{jon}, significant amounts of impurities are also clearly present as evidenced by the residual linear term in specific heat\cite{mu,prb,kim,hardy1}, $\mu$SR\cite{williams}, optical conductivity\cite{o1,o2}, the broadening of the NMR lineshape with doping\cite{ning} and the gap inhomogeneity observed by STM\cite{yin,m}. Some inhomogeneity may be unavoidable as alloying is currently required to achieve superconductivity at the highest temperatures. However, the availability of cleaner crystals could expose the intrinsic power law behaviors necessary to identify the superconducting gap structure and permit better phase sensitive experiments to be performed. In addition, the evolution in low temperature properties between clean and dirty samples can itself be a useful probe of the SC order parameter (see Ref.\onlinecite{bal} and references therein). For instance, the penetration depth of samples with varying amounts of heavy ion irradiation was recently studied\cite{h.kim}. Annealing is another typical route in the opposite direction: it improves crystal quality. Indeed, for doped SrFe$_{2}$As$_{2}$ it was recently shown that annealing can significantly reduce scattering in the normal state, as well as enhance $T_{c}$\cite{gillet,ss}.

Fe-based superconductors are widely believed to have a so-called $s\pm$ gap structure, belonging to a fully symmetric $A_{1g}$ representation but with a sign change of the order
parameter between the electron and the hole Fermi surface
sheets~\cite{Mazin2008splus,Chubukov2008rg,Seo2008splus,zlatko}. While the gap on
the hole Fermi surface (FS) sheets is believed to be nearly isotropic, there
are indications from both
theory~\cite{Chubukov2009nodes,Maier2009,Kuroki2009switch} and experiment~\cite{new,Reid,muschler,martin,ha} that on the electron sheet a strongly anisotropic, or even nodal, gap is
produced by the competition of the magnetically-assisted interband scattering
and the Coulomb interaction within each band (electron or hole).
The anisotropy is expected to be more pronounced
on the overdoped side.

Impurities are crucial for determining the precise low-energy
density of states and, hence, the nature of the superconducting order parameter. First, the impurity scattering within each band
tends to make the gap on the corresponding FS sheet more isotropic, and may
result in lifting the nodes and removal of the low-energy excitations
~\cite{Mishra2009lift,Mishra2009thcn}. In contrast, the nonmagnetic scattering by the
same impurities between bands with an opposite sign of the
superconducting order is pairbreaking, and increases the number of unpaired
quasiparticles in the gap. For impurity concentrations exceeding a critical
value (which depends on the strength of the individual scatterers and the
relative strength of the inter- vs intra-band scattering) the system possesses
low energy excitations. Therefore a study of these excitations and their
correlations with impurity strength and concentration addresses the nature of
superconductivity in pnictides.

\begin{figure*}
\begin{minipage}{0.9\textwidth}
\includegraphics[width=0.451\textwidth]{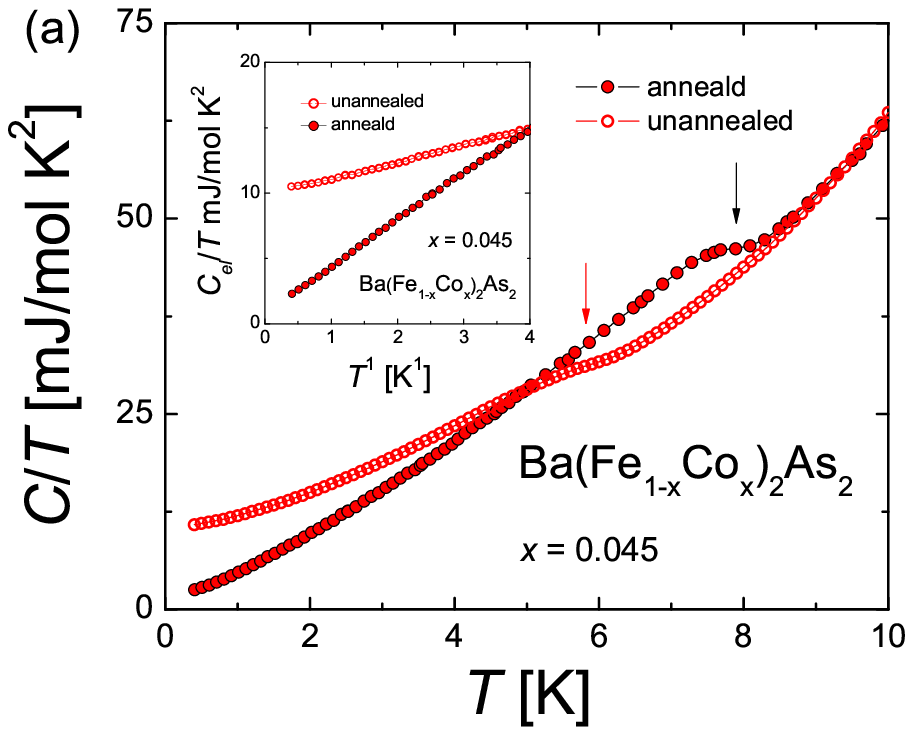}
\includegraphics[width=0.451\textwidth]{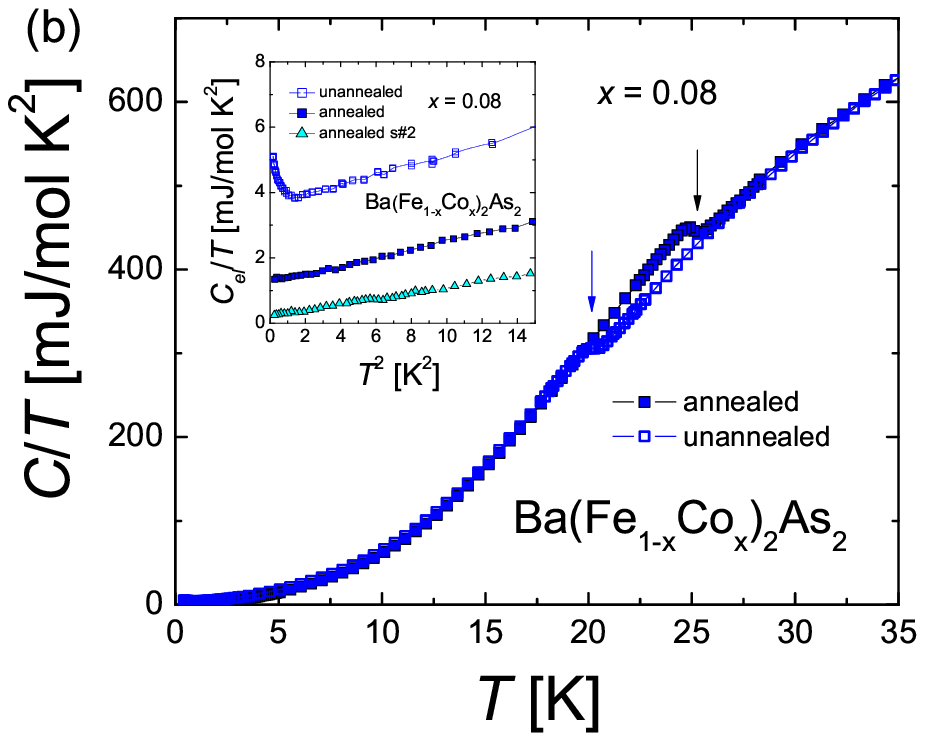}
\includegraphics[width=0.451\textwidth]{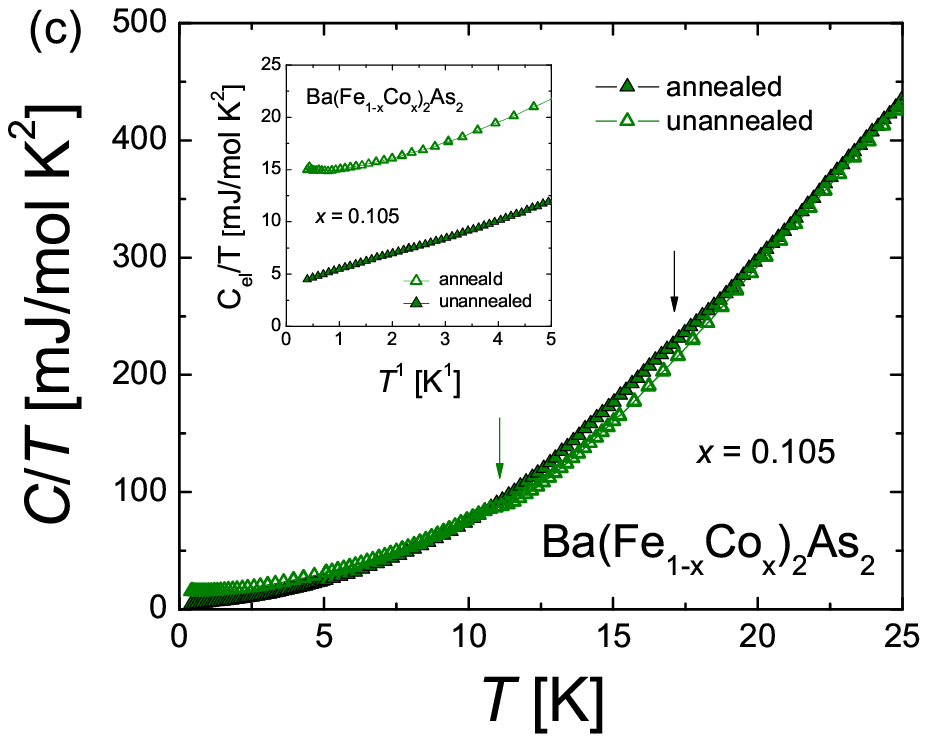}
\includegraphics[width=0.451\textwidth]{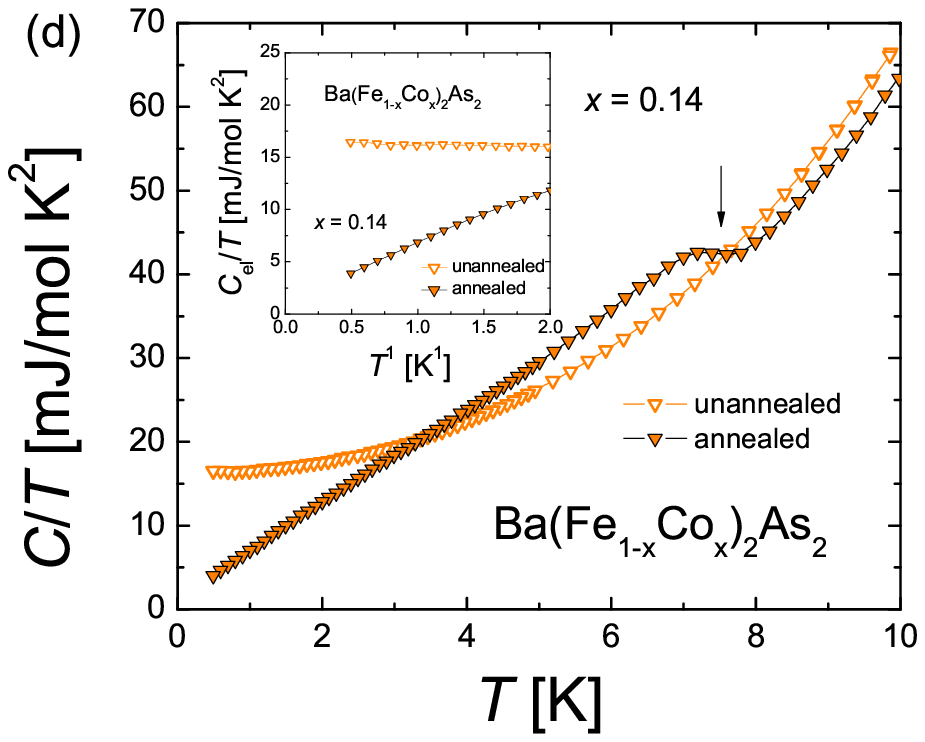}
\end{minipage}
\setlength{\unitlength}{1in}
\begin{picture}(0,0)(0,0)
\end{picture} \vspace{-0.17cm}
\caption{(Color online) The temperature dependence of the heat capacity of Ba(Fe$_{1-x}$Co$_{x}$)$_{2}$As$_{2}$ before (red circles) and after (blue squares) annealing for: (a) x = 0.045, (b) x = 0.08, (c) x = 0.105 and (d) x = 0.14. Arrows mark the superconducting transitions. Inset: low-temperature part of the electronic specific heat of Ba(Fe$_{1-x}$Co$_{x}$)$_{2}$As$_{2}$.}\label{fig1}
\end{figure*}

One way to address these states is by studying the low-temperature heat capacity. The measurements of the temperature dependence of the specific heat and its magnetic response in the superconducting state give important information about the symmetry of the order parameter\cite{hussey,FB,naka}. For near optimally Co doped BaFe$_{2}$As$_{2}$ such an analysis suggest the presence of an anisotropic gap structure which may or may not have nodes (see Ref.\onlinecite{njp,hardy1,mu,tuson}). As have been shown in the latter papers, specific heat displays significant residual density of states as a consequence of impurities of some form (see also Ref.\onlinecite{prb}). Moreover, the heat capacity at very low temperatures is very sensitive on the crystals quality and very often dominated by a Schottky-like contribution\cite{kim}. This can make the gap structure analysis quite complex.

In this paper, we present results of our specific heat studies of the under-, optimally- and over-doped Ba(Fe$_{1-x}$Co$_{x}$)$_{2}$As$_{2}$ as a function of annealing. Some preliminary results at optimal doping were presented in Ref.\onlinecite{SCES}. We show this process improves significantly superconducting characteristics in the Co-doped BaFe$_{2}$As$_{2}$ (Ba122) system. It increases $T_{c}$ and results in a pronounced decrease of the residual linear term of the specific heat. Moreover, after annealing, we do not observe a low-temperature Schottky anomaly. Subsequently, we examine the low temperature and magnetic field behavior of the electronic contribution to the specific heat and compare with self consistent calculations based on an extended $s$-wave pairing mechanism with impurities. The suppression of low energy excitations as a function of annealing in the optimally doped sample suggests a fully gapped superconducting state. In contrast, the evolution with annealing at under and over-doping indicate that the superconducting gap structure contains nodes at these doping levels. This implies a change in the gap structure with doping.

\section{Experimental details}

Single crystals were grown out of FeAs flux with a typical size of about 2$\times$1.5$\times$0.2 mm$^{3}$ (see Ref.\onlinecite{122b}). They crystallize in well-formed plates with the [001] direction perpendicular to the plane of the crystals. The doping levels were determined by microprobe analysis. After synthesis all samples have been characterized and studied, then we annealed the samples for two weeks at 800~$^{o}$C in vacuum and repeated the measurements on the same pieces. The heat capacity was measured down to 400~mK and in magnetic fields up to 9~T using a thermal relaxation method implemented in a Quantum Design PPMS-9 device. All specific heat data measured in field were field cooled.

\section{Results}

Fig.1 shows the temperature dependent specific heat of annealed and unannealed Ba(Fe$_{1-x}$Co$_{x}$)$_{2}$As$_{2}$. As may be seen from the figure the annealing process has a dramatic impact on the specific heat of the samples measured. It increases $T_{c}$ between unannealed and annealed samples quite substantially from 5.6 to 8~K (underdoped), 20 to 25~K (optimaldoped), 11 to 17.2~K (overdoped, x = 0.105) and non-superconducting to 7.45 K (over-doped x = 0.14), where $T_{c}$ was determined by equal area construction of the specific heat jump. Moreover, the annealing process strongly suppresses the residual specific heat $\gamma_{0}$ from 10.5, 3.6 and 14.6 mJ/mol~K$^{2}$ to 1.3, 1.3 and 3.8~mJ/mol~K$^{2}$ respectively for under-, optimal- and over-doped samples. In the inset of Fig.1b we have also included data for an annealed optimally doped sample from a different batch (sample s\#2). This sample has $T_{c}$~=~26~K and the residual specific heat as low as 0.25~mJ/mol~K$^{2}$ which is the lowest reported value within the Co-doped Ba122 system. Furthermore, no Schottky-like anomaly is observed for the annealed materials. In addition to the increase of $T_{c}$, the transition width broadens significantly for the overdoped sample, while it remains the same for the optimally-doped sample. For the underdoped annealed sample a reasonably sharp jump at 8~K exists in addition to a broad tail which starts at 11~K, the temperature at which the resistive transition goes to zero (see Fig.6). A similar broadening is observed in susceptibility measurements of Co-doped SrFe$_{2}$As$_{2}$\cite{gillet} and may indicate that the annealing process is not uniform throughout the sample in some instances.

\begin{figure}[t!]
\begin{centering}
\includegraphics[width=0.5\textwidth]{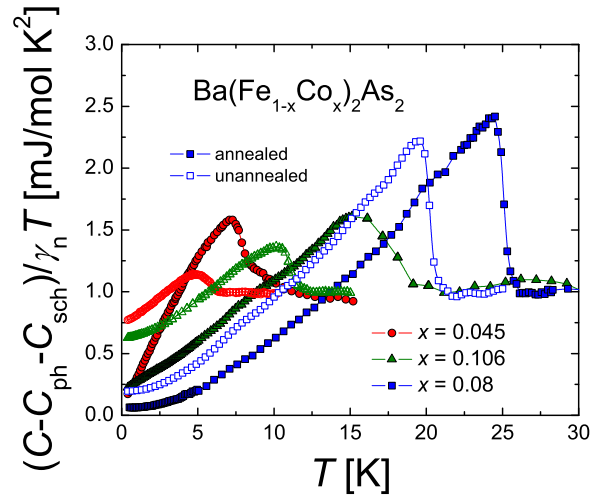}
\caption{(Color online) The temperature dependence of the electronic fraction of Ba(Fe$_{1-x}$Co$_{x}$)$_{2}$As$_{2}$ normalized by normal state specific heat (circles: x = 0.045, triangles: x = 0.105, squares: x = 0.08). Empty and full symbols represent unannealed and annealed data respectively.}\label{fig2ab}
\end{centering}
\end{figure}

As can be seen from Fig.1, above $T_{c}$ the $C(T)$ curves overlap indicating very similar temperature dependence of the phonon contributions in all these materials. To obtain the electronic part of the specific heat, we have used the same approach used previously for the optimal and doped  Ba(Fe$_{1-x}$Co$_{x}$)$_{2}$As$_{2}$ samples (see Ref.\onlinecite{njp,hardy1,prb}). We have assumed that the phonon contribution of the specific heat is independent of doping and we use the phonon specific heat obtained from the parent compound. Then we separate the lattice contribution to the specific heat of the pure BaFe$_{2}$As$_{2}$ compound as $C_{ph}(T)$~=~$C(T)^{BaFe_{2}As_{2}}$ - $\gamma_{el}^{BaFe_{2}As_{2}}T$ where $\gamma_{el}^{BaFe_{2}As_{2}}$ is the T$\rightarrow$0 intercept of $C/T$ of BaFe$_{2}$As$_{2}$. So, the electronic specific heat is simply determined by $C_{el}(T)^{Ba(Fe_{1-x}Co_{x})_{2}As_{2}}~=~C(T)^{Ba(Fe_{1-x}Co_{x})_{2}As_{2}}-C(T)_{ph}$.

\begin{figure}[t!]
\begin{centering}
\includegraphics[width=0.5\textwidth]{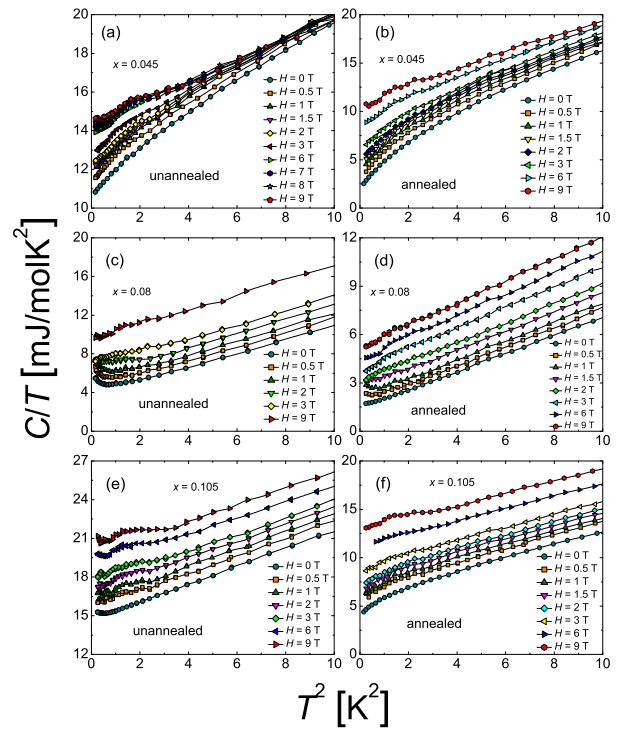}
\caption{(Color online) The low temperature dependence of the specific heat of the unannealed and annealed samples measured at different magnetic fields for H$\|$c. The data are presented in the form of $C/T$ vs. $T^{2}$ (see text)}\label{fig3}
\end{centering}
\end{figure}

\begin{figure}[t!]
\begin{centering}
\includegraphics[width=0.5\textwidth]{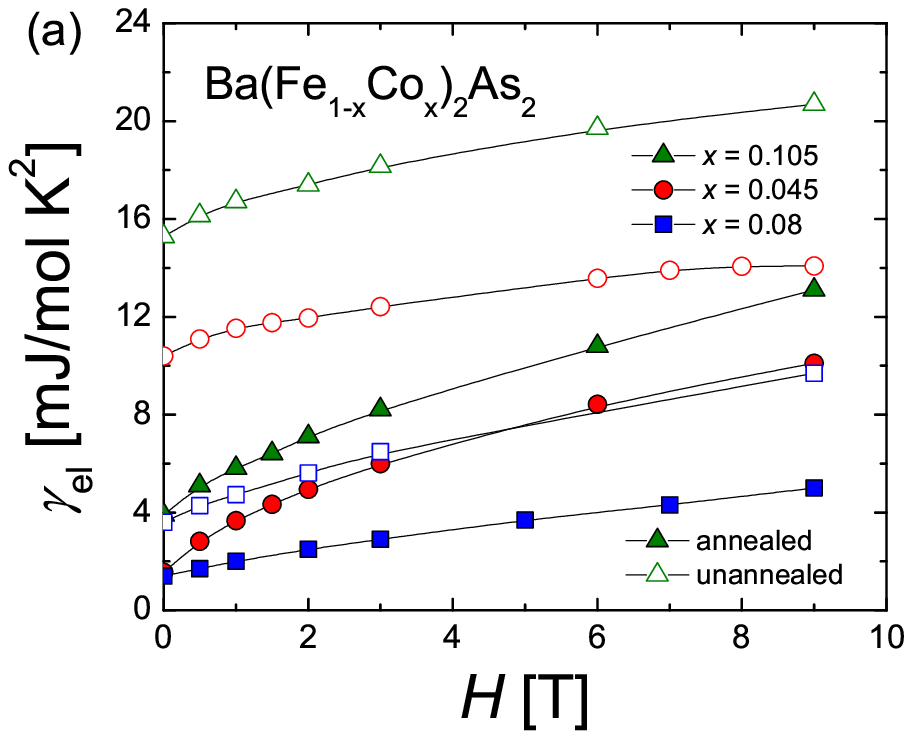}
\includegraphics[width=0.5\textwidth]{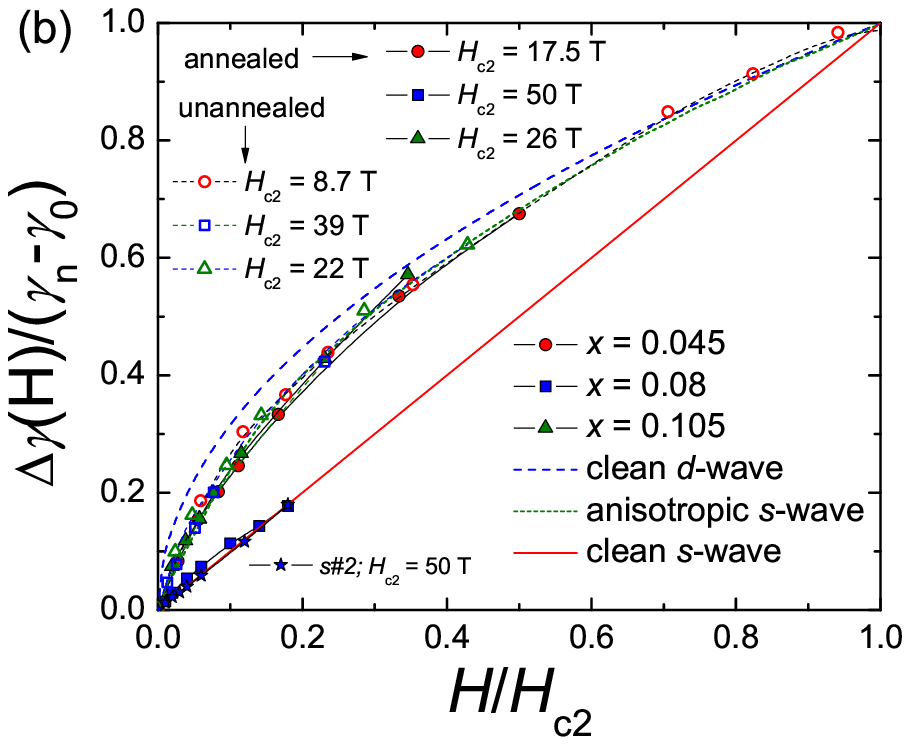}
\caption{(Color online) (a) The magnetic field dependence of the low temperature specific heat of Ba(Fe$_{1-x}$Co$_{x}$)$_{2}$As$_{2}$ (circles: x = 0.045, triangles: x = 0.105, squares: x = 0.08). Empty and full symbols represent unannealed and annealed data respectively. (b) the same data as above presented as $\Delta\gamma(H)/(\gamma_{n}-\gamma_{0})$ vs. $H/H_{c2}$. The doted, solid and dashed lines described the field dependencies of the low temperature specific heat according to clean $s$-\cite{Caroli}, anisotropic $s$-\cite{nakai} and clean $d$-wave\cite{volovik} models respectively.}
\label{fig2ab}
\end{centering}
\end{figure}

Fig.2 displays the electronic part of the specific heat of unannealed and annealed Ba(Fe$_{1-x}$Co$_{x}$)$_{2}$As$_{2}$. In both cases it is obtained by subtracting the phonon contribution, together with a small Schottky-term below 1~K (for the unannealed sample), and normalizing by $\gamma_{n}$. A pronounced jump associated with the superconducting transition can be observed. As has been shown previously, none of the specific heat data can be described by a single $s$ or $d$-wave model\cite{njp,prb,hardy1,zeng}. This is also true for other Fe-based superconductors (see for instance Ref.\onlinecite{pop, kim2, mu2}).
Clearly there are additional low energy excitations not captured by a single gap and, at least, a two gap scenario is necessary to describe the experimental results in this system\cite{njp,prb,hardy1,zeng}. Furthermore, several curves have a hump at intermediate temperatures (for instance the annealed overdoped sample at 9~K) which can be taken as evidence for multiband SC as in MgB$_{2}$\cite{FB}. In addition to the suppression of the residual linear and Schottky terms, the finite temperature dependence of $C_{el}/T$ for underdoped and overdoped samples changes with annealing, while there is no significant change at optimal doping. This can be seen clearly in the insets of figure 1. In particular, the low temperature dependence of the overdoped sample changed from $C_{el}/T=\gamma_{0}$+$BT^{1.7}$ for the unannealed crystal to $C_{el}/T=\gamma_{0}$+$BT^{1}$ for the annealed crystal. For the underdoped annealed sample the temperature range over which the behavior $C/T=\gamma_{0}$+$BT^{1}$ is visible is limited due to the lower $T_{c}$, but the change in the temperature dependence at finite temperature is readily apparent from figure 2. In none of the samples do we find evidence for the $C\sim T^{3/2}$ behavior expected for samples with accidental points nodes in 2D\cite{zt} perhaps due to the finite number of impurities present in even the cleanest sample measured.

The temperature dependence of the specific heat of annealed and unannealed samples, taken at different magnetic fields applied along the c-axis is shown in Fig.3. The field dependence of the residual linear term plotted in fig.4a is extracted from this data by extrapolating the low temperature specific heat to 0~K (from the temperature above the Schottky anomaly). $\Delta\gamma(H)\equiv\gamma_{0}(H)-\gamma_{0}(0T)$ reflects the rate of quasiparticle production with magnetic field. For the unannealed optimally doped and overdoped samples (Figs 3c and 3e), aside from a broadening of the low temperature upturn with increasing field, $C/T$ shifts rigidly upward as has been reported previously\cite{njp,prb,kim}. The effect of annealing has remarkably different effects on the field dependence of the two dopings however. For optimal doping the effect of a magnetic field on the annealed crystal continues to be a rigid shift of $C/T$ with increasing magnetic field, although the magnitude of $\Delta\gamma(H)$ is \textit{reduced} by a factor of two relative to the unannealed crystal. For the overdoped sample however, the change in the residual linear term as a function of magnetic field $\Delta\gamma(H)$ \textit{grows} by 40$\%$ in the annealed crystal compared with the unannealed sample. Furthermore, as we noted above the overdoped annealed sample has a temperature dependence of $C/T$=$\gamma_{0}$+$BT^{1}$ in the low temperature limit. The application of a magnetic field restores the temperature dependence to $C/T$=$\gamma_{0}(H)+BT^{2}$. The behavior on the underdoped crystals is qualitatively similar to that for the overdoped crystals. Again annealing enhances $\Delta\gamma(H)$ from 3.7 to 8.8~mJ/mol~K$^{2}$ in the underdoped sample.

In an effort to determine whether the opposite evolution in the field dependence as a function of annealing was due to a simple change in $H_{c2}$ or [$\gamma_{n}$-$\gamma_{0}$(0T)] we plot in figure 4b the field dependence of the residual linear term of the specific heat scaled by [$\gamma_{n}$-$\gamma_{0}$(0T)] versus $H/H_{c2}$. $H_{c2}$ was obtained from the specific heat determined $T_{c}(H)$ and applying the WHH formula\cite{WHH} to obtain $H_{c2}$(0), and is in reasonable agreement with the measured values of $H_{c2}$\cite{ni} where available. For under and over-doping the field dependence of $\gamma$ scales with the unannealed crystals. However, at optimal doping there is a dramatic departure from this scaling. The field dependence resembles the expectation for an isotropic $s$-wave superconductor, where the number of Caroli-deGennes bound states increases linearly with field due to the linear increase in the number of vortices entering the sample\cite{Caroli}. Alternatively, it would require an upper critical field of ~125~T to scale the data from the optimally doped annealed crystal with the remaining curves. This behavior is even more prominent in the annealed optimally doped sample with $\gamma_{0}$=0.25~mJ/mol~K$^{2}$ (sample s\#2) where the data perfectly follows the prediction for an isotropic $s$-wave superconductor (see the filled stars in Fig.4b). Consequently there exists a qualitative difference as to the effects of annealing between optimally doped and non-optimally doped crystals. A table summarizing the features described above are shown in Table I.

\begin{table*}
\caption{\label{tab:table3}Parameters obtained by fitting the equation $C_{el}/T=\gamma_{0}+BT^{\alpha}$ to the experimental data (below 5~K) of under (x=0.045), optimal (x=0.08) and overdoped (x=0.105 and 0.14) Ba(Fe$_{1-x}$Co$_{x}$)$_{2}$As$_{2}$. Arrows mark the change (increase or decrease) of the parameter for the annealed sample compared to the unanealed one.}
\begin{ruledtabular}
\begin{tabular}{ccccccc}
sample [x]&$T_{c}$ [K]&$\gamma_{0}$ [mJ/molK$^{2}$]&Schottky
&$B$ [mJ/molK$^{\alpha+2}$]&$\alpha$&$\Delta\gamma(9T)$\\ \hline
0.045 unannealed&5.6&10.2&No& 0.97 & 1.2 & 3.7  \\
0.045 annealed&8.0 $\uparrow$&1.0 $\downarrow$&No& 3.71 $\uparrow$& 0.98 $\downarrow$& 8.8 $\uparrow$\\
\hline
0.08 unannealed&20&3.6&Yes& 0.12 & 2.1 & 6.07\\
0.08 annealed&25 $\uparrow$&1.3 $\downarrow$&No $\downarrow$& 0.09 $\downarrow$& 2.2 $\uparrow$& 3.7 $\downarrow$\\
0.08 annealed s\#2&26 $\uparrow$&0.25 $\downarrow$&No $\downarrow$& 0.04 $\downarrow$& 2.4 $\uparrow$& 3.85 $\downarrow$\\
\hline
0.105 unannealed&11&14.5&Yes& 0.47 & 1.7 & 6.1 \\
0.105 annealed&17.2 $\uparrow$&3.8 $\downarrow$&No $\downarrow$& 1.37 $\uparrow$& 1.05 $\downarrow$& 9.3 $\uparrow$\\
\hline
0.14 unannealed&--&--&No& -- & -- & -- \\
0.14 annealed&7.45 $\uparrow$&0.4 &No & 6.5& 0.9 & 14.2 \\

\end{tabular}
\end{ruledtabular}
\end{table*}

\section{Discussion}

The improved crystal quality of the annealed crystals as evidenced by the increase in $T_{c}$, reduction in the residual linear term of the specific heat, and suppression of any low temperature Schottky upturn puts us in a better position to try to understand the order parameter symmetry of Co-doped BaFe$_{2}$As$_{2}$. We first discuss several aspects of the data on a qualitative level, and subsequently we compare self-consistent calculations using a general two-band model which capture many of the salient features of the temperature dependent specific heat data at optimal and overdoping.

\subsection{Gap phenomenology}

For the annealed optimally doped sample, where the residual linear term is the smallest, the zero field specific heat of the annealed sample has a long temperature limiting behavior of $C_{el}/T$=$\gamma_{0}$+$BT^{2}$ (see inset of Fig.1). The value of $B$ is 25~$\%$ larger than the $T^{2}$ term coming from the acoustic phonon branch. It is very difficult to get this precise low temperature dependence from a simple gap structure. Clean point nodes are capable of producing a specific heat $C\sim T^{3}$, however, any impurities which would be needed to provide the finite $\gamma_{0}$ term should also lift the point node and consequently remove the expected $T^{3}$ dependence. The fit which spans nearly a decade in temperature argues against this simply being a crossover regime, although we can not rule it out entirely. Finally, bosonic modes might easily produce a $C\sim T^{3}$ contribution to the specific heat. Perhaps due to the suppression of the magnetic and/or structural transition one may find such bosonic excitations which were not present in or modified from the parent\cite{njp} or heavily overdoped compounds\cite{hardy1}. Regardless of the unknown origin of the $T^{3}$ term at optimal doping, it is clear that in approaching the clean limit there are very few low energy electronic excitations. This is supported not only by the smallest residual linear term and the small increase in $C/T$ with increasing temperature, but also by the fact that $\Delta\gamma$(9T) is the smallest in the optimally doped annealed sample. In fact, the normalized field dependence shown in figure 4b would indicate a nearly fully gapped superconductor.

At optimal doping, prior to annealing, $T_{c}$ was lower indicating the existence of more impurities. In the unannealed sample $C/T$ has a larger residual $\gamma$, and there also exists a larger field dependence to $\gamma$ relative to the annealed sample. These statements are consistent with a decrease of low energy excitations with annealing. A pure phase separation scenario can not explain both of these observations. Rather a gap structure which creates impurity states at low energies may be able to reproduce such an effect.\\

The situation changes dramatically when moving to either under or over doping. We discuss the situation in going from optimal doping to overdoping, and note that everything is qualitatively similar in going to underdoping as well. The annealed overdoped sample relative to the annealed optimally doped sample has several differences. The temperature dependence of the electronic specific heat for the overdoped samples is much better expressed as $C/T$=$\gamma_{0}+BT^{1}$ (as opposed to $C/T$=$\gamma_{0}$+$BT^{2}$) in which the magnitude of both terms is larger than that found at optimal doping. In addition, $\Delta\gamma(H)$ is larger, and is clearly sublinear. Furthermore, the temperature dependence evolves from $C/T$~=~$\gamma_{0}+BT^{1}$ at zero field to $C/T$~=~$\gamma_{0}$+$BT^{2}$ at high fields. All of these features of the annealed over-doped sample are consistent with a gap structure that either has nodes, or very deep minima. Similar conclusions were previously found form a high quality overdoped crystals grown by a different technique\cite{tuson}.

The similarities and differences in the evolution of specific heat features with annealing
for the overdoped (underdoped) and the optimally doped materials is also instructive (see table I).
With annealing $T_{c}$ increases, $\gamma_{0}$ drops, and a low
temperature Schottky term disappears in all materials.
On the other hand, in both overdoped and underdoped cases, the $BT^{\alpha}$ and $\Delta\gamma(H)$ terms
are enhanced in cleaner (annealed) samples, indicating
that there are clearly more electronic excitations associated with the gap structure in the annealed case. This is precisely opposite to our conclusion at optimal doping. The situation for the underdoped crystal is qualitatively identical to the overdoped crystal. We cannot rule out the possibility that further improvement in sample quality for the overdoped and underdoped samples may lead to temperature and field dependencies which resemble the optimally doped sample. However, the qualitatively different trends with annealing suggests that the gap structure must qualitatively change as a function of doping.

\subsection{Comparison to theory}

Currently the most highly discussed model for the superconducting gap in the iron pnictides is the so called $s^{\pm}$ state which receives theoretical support from spin fluctuation models of various tight binding parameterizations of the Fermi surfaces of these materials\cite{Mazin2008splus,Chubukov2008rg,Seo2008splus,zlatko}.

\begin{figure}[t!]
\includegraphics[width=\linewidth]{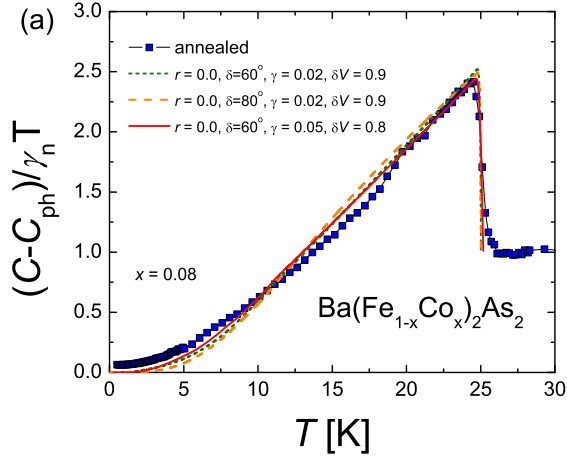}
\includegraphics[width=\linewidth]{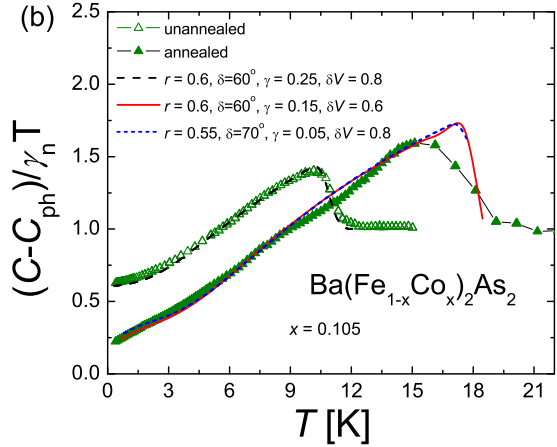}
\caption{(Color online) (a) Optimal doping, $x=0.08$.
	Theoretical fits (lines) to experimental data for annealed sample. (b) Overdoped regime, $x=0.105$.
	Theoretical fits (lines) to experimental data for unannealed (open symbols) and annealed (filled symbols) sample.}
\label{figODfits}
\end{figure}

We computed the heat capacity within a model of equal size
electron and hole Fermi surfaces with an interband pairing that
supports an isotropic gap on the hole FS sheet, $\Delta_h(\phi)=\Delta_h$
and an anisotropic
gap of the general form $\Delta_e(\phi)=\Delta_e (1-r +r \cos 2\phi)$
on the electron sheet, with parameter $r$ varying from $r=0$ (isotropic $s^\pm$ state)
to $r=1$ (equal-size positive and negative lobes). The ratio of $\Delta_e$ to $\Delta_h$ is determined by our choice of equal density of states on the hole and electron Fermi surfaces. We accounted for the
impurity scattering in the self-consistent T-matrix approximation, and
solved the resulting gap equation to obtain the density of states and
the entropy as a function of temperature. The impurity scattering is qualified by $\tilde\gamma=\frac{\Gamma}{2\pi T_{c0}}$ where $\Gamma=\frac{n_{imp}}{\pi N(E_{f})}$ and $n_{imp}$ is the impurity concentration.

It is clear from the data that the optimally doped sample exhibits
significantly smaller temperature variation than either over- or under-doped
materials, suggesting a very small density of low-energy excitations. Consistent
with that observation, we find the best fit for these data is using a fully
isotropic gap on the electron Fermi surface, $r=0$, with
relatively few impurities of moderate strength given by the phase shift $\delta=60-80^\circ$
but significant interband scattering $\delta V = v_{12}/v_{11}=0.8-0.9$,
Fig.~\ref{figODfits}a. For $r$~=~0.0, $\delta$~=~60$^\circ$, $\gamma$~=~0.05 and $\delta V$~=~0.8, the calculation gives $\Delta_e$~=~-$\Delta_h$~=~3.8~meV. On the overdoped side, we found that several possible parameter values
give an adequate fit to the data at the intermediate temperatures. However, the
low-temperature behavior is most consistent with a nodal order parameter on
the electron Fermi surface sheet ($r>0.5$). For fully gapped systems, even when
the impurity concentration is sufficient to produce a non-vanishing density of
states at zero energy, we find significantly more features in the calculated
$C/T$ at low and intermediate temperatures, due to the strong variation of the
DOS between the upper edge of the impurity band and the superconducting gap
edge\cite{Mishra2009thcn}. In contrast, in nodal systems the evolution of the Sommerfeld coefficient
is smooth as shown in Fig.~\ref{figODfits}b, and the choices of $r=0.55$ or
$0.6$, with closely spaced nodes on the electron FS sheets, give the best
agreement with experiment. For $r$~=~0.6, $\delta$~=~60$^\circ$, $\gamma$~=~0.15 and $\delta V$~=~0.6, the calculation gives $\Delta_e$~=~4~meV and $\Delta_h$~=~-2.4~meV. The linear behavior down to the lowest temperatures
suggests, from our fits, scatterers of intermediate strength, phase shift of
close to 60$^\circ$, and a substantial ratio of inter- to intraband scattering.
This is plausible in Co-doped systems where the dopants are on the Fe
sites and directly affect the $d$-orbitals contributing to both FS sheets. Using
the same phase shift, but increasing the impurity concentration and slightly
increasing the relative strength of the interband scattering, we are able to
fit well the unannealed data as well, as shown in Fig.~\ref{figODfits}b. We also roughly account for the 64~$\%$ suppression of $T_{c}$ (17.2 to 11~K) by increasing the impurity concentration from 0.1 to 0.25 using $r$~=~0.6, $\delta$~=~60$^{o}$ and $\delta V$~=~0.6 gives a 62~$\%$ suppression of $T_{c}$ (from 0.77 $T_{c0}$ to 0.48 $T_{c0}$).
The underdoped sample is in the region of the coexistence of magnetism and superconductivity. Consequently, we did not attempt a fit since our theory does not include the effects of the magnetic ordering that is likely significant in this regime.

\subsection{Effects of annealing}

Annealing clearly improves sample quality, as $T_{c}$ increases, the residual linear term decreases, and the low temperature Schottky anomaly is suppressed; however, the microscopic origin for these effects is less clear. The change of $T_{c}$'s could suggest a change of cobalt concentration in the samples after annealing resulting in a shift in doping towards higher $T_{c}$ on the phase diagram. However, this could be ruled out by the fact that $T_{c}$ increases for both the under-doped and overdoped samples. Moreover, in the underdoped material, the spin density wave and structural transitions also increased (see Fig.6) which would contradict an increase in doping to account for the increase in $T_{c}$. The increase in the spin density wave transition with annealing is also observed in the parent compound\cite{rotundu} and in Co-doped SrFe$_{2}$As$_{2}$\cite{gillet}.

\begin{figure}[ht]
\begin{centering}
\includegraphics[width=0.5\textwidth]{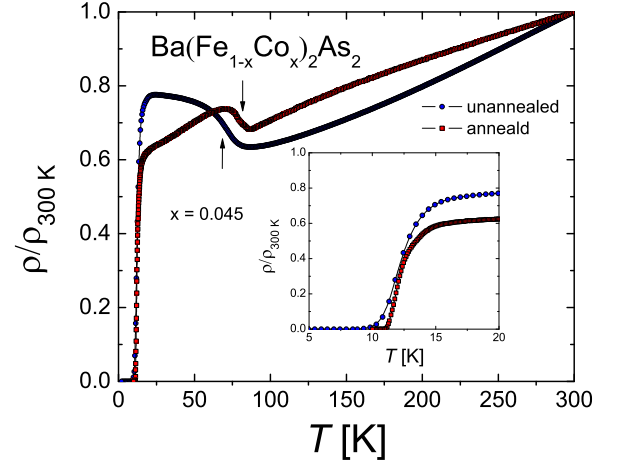}
\caption{(Color online) The temperature dependencies of normalized electrical resistivity of unannealed (squares) and annealed (circles) Ba(Fe$_{0.955}$Co$_{0.045}$)$_{2}$As$_{2}$. Arrows mark the antiferromagnetic phase transition. Inset: low temperature part of the resistivity.}\label{fig6}
\end{centering}
\end{figure}

\begin{figure}[ht]
\begin{centering}
\includegraphics[width=0.5\textwidth]{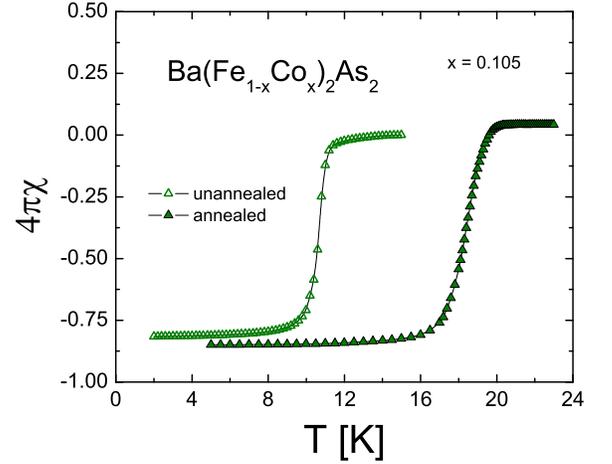}
\caption{(Color online) The temperature dependencies of magnetic susceptibility of unannealed (empty symbols) and annealed (filled symbols) Ba(Fe$_{0.895}$Co$_{0.105}$)$_{2}$As$_{2}$. The ZFC measurements have been performed on the same sample with the filed parallel to the ab plane of the crystal.}\label{fig7}
\end{centering}
\end{figure}

The presence of a finite residual linear term could originate from a sample which contains both superconducting and non-superconducting portions of a sample. Consequently, one might imagine that annealing is simply trading non-superconducting portions for superconducting portions. However, as shown in Fig.7, the volume of the superconducting fraction, as measured by magnetic susceptibility, does not change significatively for the x = 0.105 sample. The fact that the overall magnetic field dependence at optimal doping is suppressed in the annealed sample further disproves such a suggestion\cite{prb}. Some fraction of the residual gamma may still originate from non-superconducting portions of the sample, but we believe the change as a function of annealing is most likely accounted for by a difference in the number of pair breaking impurity scatterers. However, we should caution that annealing clearly does more than just change the number of pair breaking impurity sites, as evidenced by its effect on the structural and SDW transitions. Also, the low temperature Schottky term is likely a result of free paramagnetic Fe or Co moments within the sample. Its suppression as a function of annealing indicates that these sites have either been removed from the sample, or possibly incorporated into the lattice. It is worth noting that a similar Schottky-type anomaly was also observed in the cuprates but, up to now, the orgin of this effect remains unknown. Furthermore, the normal state Sommerfeld coefficient also changes with annealing from 13.7, 18, and 23.2 mJ/mol~K$^{2}$ prior to annealing to 14, 22, and 20 mJ/mol~K$^{2}$ after annealing for the under-, optimally-, and over-doped samples, respectively.

While the temperature and field dependencies for the annealed samples now allow us to identify some difference in low energy gap structure as a function of doping, it is interesting that the temperature and field dependence of all dopings with significant impurity concentrations could be scaled (after subtraction of the residual linear term) with one another\cite{prb}. This scaling persists in the temperature dependence of the annealed optimally doped sample\cite{SCES}, but not its field dependence. However, the temperature dependence of the annealed under and overdoped samples no longer scale with the previous sets of data.

All the results obtained clearly indicate that more work is needed to understand the microscopic effects of annealing, and whether further improvement in sample quality can help elucidate additional details of the pairing mechanism.

\section{Summary and conclusions}

In summary, we report on the effect of annealing on the specific heat of under-, optimally- and over-doped
Ba(Fe$_{1-x}$Co$_{x}$)$_{2}$As$_{2}$. We have shown that annealing at 800$^{o}$C for two weeks improves superconducting characteristics in the Co-doped BaFe$_{2}$As$_{2}$ system significantly. The heat treatment results in an increase of the critical temperatures and largely decreases the residual specific heat in all samples. Also, no Schottky-like term is observed at low temperatures.

From the temperature and magnetic field dependence of the electronic specific heat of the annealed crystals we find a significant reduction of thermally excited low energy excitations for optimal doping relative to the under and over-doped regimes. This suggests a change in the gap structure as a function of doping which is consistent with the lifting of accidental nodes at optimal doping. Similar behavior has also been seen in Zn doping studies\cite{new}, thermal transport\cite{Reid}, Raman scattering\cite{muschler}, $\mu$SR\cite{williams} and penetration depth measurements\cite{martin}. Whether the current theoretical models can self consistently explain all these experimental observations remains to be seen.

We note that the quality of samples has a strong influence on the power law observed in the low temperature specific heat data. Furthermore, the low temperature Schottky anomaly is also sample dependent and will likely manifest itself differently for different techniques. Consequently, the conclusions drawn by different techniques, and even identical techniques on different samples can dramatically vary. The power laws in transport and thermodynamic measurements are heavily relied upon to determine plausible gap structures and subsequently pairing mechanisms. It is also very difficult to have a single independent measure (such as RRR) to determine the quality of a sample. Consequently, this work emphasizes the need for multiple measurements to be carried out on the same high quality sample where possible.

\begin{acknowledgments}

We gratefully acknowledge fruitful discussions with
A. V. Balatsky, T. Park and G. Stewart. Work at Los Alamos National Laboratory was
performed under the auspices of the U.S. Department of
Energy, Office of Science and supported in part by the Los Alamos LDRD program. Research at ORNL is sponsored by the Materials Sciences and Engineering Division, Office of Basic Energy Sciences, U.S. Department of Energy. A.S acknowledges discussions with M. A. McGuire, D. Mandrus and B. C. Sales. I.V. acknowledges support from DOE Grant DE-FG02-08ER46492. A.B.V acknowledges support from NSF grant DMR-0954342. The work at McMaster was supported by NSERC, CFI, and CIFAR
\end{acknowledgments}

\end{document}